\documentclass[12pt]{article}
\usepackage{esfconf}
\begin{document}

\title{Singular potentials and absorption problem in Quantum Mechanics}
\authors{Vladimir D. Skarzhinsky\adref{1,2} and J{\"u}rgen Audretsch\adref{2}}
\addresses{\1ad P.~N.~Lebedev Physical Institute, Leninsky prospect 53,
Moscow 117924, Russia,
\nextaddress
\2ad Fakult\"at f\"ur Physik der Universit\"at Konstanz,
Postfach M 673,\\ D-78457 Konstanz, Germany}

\maketitle

\begin{abstract}
We discuss a possible approach to the absorption problem in Quantum
Mechanics based on using of singular attractive potentials in the
corresponding Schr\"{o}dinger equations. Possible criteria for selection
of exact solutions of these equations are considered and it is shown
that different models of absorption can be realized by a special choice
of exact solutions. As an example, the motion of charged particles in
the  Aharonov-Bohm (AB) and the scalar attractive $\rho^{-2}$ potentials
is investigated in detail. Other attractive potentials are briefly
considered.
\end{abstract}

\section{Introduction}

The introduction of complex potentials is generally accepted way to
describe absorption at quantum scattering \cite{Hodgson63}. Based on an
analogy with electrodynamics this method can be, nevertheless, disputed
in QM because of quite different nature of wave functions and
electromagnetic field. Proposed recently approach to the absorption using
restricted Feynman-type integrals \cite{March99} seems to be more adequate
to this problem. Also, absorption can be treated in connection with
the well-known problem of collapse of particles scattered by singular
potentials \cite{Landau77}. In contrast to  the case of regular potentials
the scattering by singular potentials leads to two linear independent
solutions of radial Schr\"{o}dinger equations that are both
square-integrable and form overfull nonorthogonal set. This means that
corresponding Hamiltonians are not self-adjoint. To describe elastic
scattering the procedure of self-adjoint extension \cite{Akhiezer93} has
been applied by imposing suitable boundary conditions at singularity
points. For the case of the singular attractive potential $\sim r^{-n},
n\geq 2$ they have been found in \cite{Case50}.

On the contrary, to describe absorption one has to refuse from the
self-adjiontness and permit non-hermitean Hamiltonians. They have been
first employed with this aim in \cite{Vogt54} where the capture of ions
by polarization forces in $\sim r^{-4}$-potential was considered. Such
interpretation of the collapse problem has been developed in
\cite{Perelom70,Alliluev72} for the case of the singular attractive
potentials $\sim r^{-n}\,, n\geq 2.$

The absorption problem became practically important in last years in
connection with atom interferometer experiments on scattering neutral
polarizable atoms by a charge wire \cite{Novak98} as well as with
superimposed magnetic field  with the aim to observe the effect of
the AB topological phase on this process. These processes are effectively
governed by the Schr\"{o}dinger equation with the AB - and the attractive
$\rho^{-2}$ potential in two dimensions. This unexpected application of
the AB effect \cite{Aharonov59} in atomic physics \cite{Wilkens94,Wei95}
has been recently discussed theoretically
\cite{Audretsch98,Audretsch99,Leonhardt98}. Absorption of atoms was an
essential part of these investigations, and the corresponding atom
interferometer experiment is on its way.

\section{Absorption and inelastic scattering in the singular AB- and
$\rho^{-2}$ potentials}

The radial wave function for scattered particles with mass $M$ and charge
$e$ in the AB-potential $A_\varphi = \beta/\rho,$ $\beta$ being the
magnetic flux in the m.f.q.units, $0\leq\beta<1,$ and in a scalar potential
$U = -\kappa^2/\rho^2$
satisfies the Bessel equation
\begin{equation}\label{Re}
R''_m (\rho)+{1\over\rho}\,R'_m(\rho)-{\nu^2\over\rho^2}\,R_m(\rho)
+p^2\,R_m(\rho) = 0
\end{equation}\,,
\begin{equation}\label{nu}
\nu^2 := (m-\beta)^2-\gamma^2\,, \quad \gamma^2 =: 2 M \kappa^2\,.
\end{equation}
The parameter $\nu^2$ can be negative.

The differential operator Eq. (\ref{Re}) is {\sl symmetric} but is not
{\sl self-adjoint}. Its self-adjoint extension depends on infinite number
of {\sl open parameters} \cite{new}.
Unfortunately, it has been not answered there of what type of physical
situations corresponds in some limit to given values of open parameters.
This turned out to be possible only for the pure AB case \cite{Audretsch95}.

The partial scattering amplitudes $f_m$ are expressed \cite{Audretsch99}
in terms of phase shifts $\delta_m$
\begin{equation} \label{f}
f_m = {e^{-i{\pi\over 4}}\over\sqrt{p}}\left(S_m-\cos\beta\right)\,,
\quad S_m := e^{2i\delta_m}\,.
\end{equation}
which are defined by the asymptotic form of the radial functions
\begin{equation} \label{asymR}
R_m(\rho) \rightarrow \sqrt{1\over 2\pi p\rho}
\left[e^{-i(p\rho-\pi m-{\pi\over 4})}+S_m e^{i(p\rho-{\pi\over 4})}\right]\,.
\end{equation}
The general normalizable solutions of Eq. (\ref{Re}) have the form
\begin{equation} \label{rs1}
R_m(\rho) = c_m J_{\mu}(p\rho)\,, \quad |m-\beta|> \sqrt{1+\gamma^2}\,,
\end{equation}
\begin{equation} \label{rs2}
R_m(\rho) = a_m H^{(1)}_{\mu}(p\rho) + b_m H^{(2)}_{\mu}(p\rho)\,,
\gamma<|m-\beta|< \sqrt{1+\gamma^2}\,,
\end{equation}
\begin{equation} \label{rs3}
R_m(\rho) = a_m H^{(1)}_{i\mu}(p\rho) + b_m H^{(2)}_{i\mu}(p\rho)\,,
|m-\beta|< \gamma
\end{equation}
where $\mu:=|\nu|$ and $J_{\mu}(x), H^{(1, 2)}_{\alpha}(x)$ are the Bessel
and Hankel functions with arbitrary coefficients $a_m, b_m.$

One can see that the normalizable wave functions can be singular at
$\rho=0$ what is typical for singular potentials. This happens for small
orbital momenta $m$ when particles spend much time nearby the singular line
$\rho=0$ and can fall onto it acquiring an infinite negative energy.
Moreover, the linear independent solutions $H^{(1)}_{\alpha}(p\rho)$ and
$H^{( 2)}_{\alpha} (p\rho)$ are both square-integrable, and we have no
criteria for selection of an unique solution to remove this ambiguity.
In what followed we discuss different possible criteria for such choice.

\section{The criteria for the selection of the unique solution}

We consider now the possible criteria for fixing coefficients of
Eqs. (\ref{rs2}), (\ref{rs3}). The behaviour of the solutions nearby the
singularity $\rho=0$ will control this procedure.
To expose what happens in general case we calculate the partial radial
currents for the solutions (\ref{rs1}) - (\ref{rs3}).
For Eq. (\ref{rs1}) the partial radial currents are equal to zero.
This means that the partial modes with $|m-\beta|> \sqrt{1+\gamma^2}$ are
always scattered elastically. Using the properties of the Hankel functions
we get for the remaining modes
\begin{equation} \label{pc2}
j_m (\rho) = {2\over \pi M\rho}\left(|a_m|^2 - |b_m|^2\right) \,,
\gamma<|m-\beta|< \sqrt{1+\gamma^2}
\end{equation}
and
\begin{equation} \label{pc3}
j_m (\rho) = {2\over \pi M\rho}\left(|a_m|^2\,e^{\pi\mu} -
|b_m|^2\,e^{-\pi\mu}\right) \,,\quad |m-\beta|< \gamma\,.
\end{equation}
The fact that the partial currents (\ref{pc2}) and (\ref{pc3}) through
a cylindrical surface of a fixed radius $\rho_0$, $2\pi\rho_0 j_m(\rho_0)$
are independent on $\rho_0$ means that the possible choice of the
coefficients and the interpretation of the corresponding solutions
have been connected with their behaviour at $\rho\rightarrow 0.$

These solutions have the following asymptotic behaviour at
$\rho\rightarrow 0$
\begin{equation} \label{asrs2}
R_m \rightarrow A_m \rho^{-\mu}+B_m \rho^{\mu}
\,, \quad \gamma<|m-\beta|< \sqrt{1+\gamma^2} \,
\end{equation}
\begin{equation} \label{asrs3}
R_m(\rho) \rightarrow A_m \rho^{-i\mu}+B_m \rho^{i\mu}\,,
\quad |m-\beta|< \gamma
\end{equation}
where the coefficients $A_m$ and $B_M$ are connected with $a_m$ and $b_m.$
For the elastic scattering the ingoing and outgoing currents have to
compensate to each other as this takes place for the modes of
Eq. (\ref{rs1}). This means that conditions  $|a_m|=|b_m|$ and
$|a_m|\,e^{\pi\mu/2}=|b_m|\,e^{-\pi\mu/2}$ must be fulfilled for
Eq. (\ref{rs2}) and Eq. (\ref{rs3}), correspondingly. This leads to the
following asymptotic behaviour of these solutions \cite{new}
\begin{equation}\label{asrs2'}
R_m(\rho)\rightarrow \rho^\mu + l_m\rho^{-\mu}\,,\quad l_m \in R_1
\end{equation}
and
\begin{equation}\label{asrs3'}
R_m(\rho)\rightarrow \rho^{i\mu} + e^{i\vartheta_m}\rho^{-i\mu}\,,
\quad 0\leq\vartheta_m<2\pi
\end{equation}
that describe the elastic scattering without absorption. These asymptotic
expression show that the solution Eq. (\ref{rs3}) contains at small
$\rho$ the ingoing and outgoing waves $e^{\mp i\mu\ln\rho}$, and the
situation looks as if there exists an additional repulsive force or a
reflective barrier with the support on $\rho=0$ which creates the outgoing
wave. The self-adjoint extension procedure admits this interpretation.
A rudiment of  this phenomenon is present in Eq. (\ref{rs2}).

Otherwise, if the conditions for the elastic scattering are not fulfilled,
absorption of particles happens on the infinitely thin wire, and it was
treated from different points of view in
\cite{Wei95,Audretsch98,Leonhardt98,Denschlag97}. The case of the total
absorption has been also considered phenomenologically in
\cite{Audretsch98,Audretsch99}. We notice that absorption can happen even
in the pure AB case, see below.

An absorption criterion has been proposed in \cite{Vogt54}, and it
has been used for the case of the attractive $\rho^{-2}$ potential for
$\beta=0$ in \cite{Perelom70,Alliluev72}. The authors have eliminated of the
outgoing wave at $\rho= 0,$ interpreting of the singularity $\rho=0$ \
as a sink. Taking $b_m\,e^{-\pi\mu} = a_m\,e^{\pi\mu}$ in Eq. (\ref{rs3}) and
keeping only the ingoing wave $e^{-i\mu\ln p\rho}$ in Eq. (\ref{asrs3}) at
$\rho\rightarrow 0$, we obtain the solution \cite{Alliluev72}
\begin{equation} \label{rs3'}
R_m(\rho) = c_m J_{-i\mu}(p\rho)\,, \quad |m-\beta|< \gamma
\end{equation}
This solution has been used in \cite{Audretsch98,Leonhardt98} to
describe the inelastic scattering of neutral polarizable atoms from the
charged wire placed in the uniform magnetic field.

In this case the partial absorption cross sections are equal to
\begin{equation} \label{abscs3}
\sigma_m^{\rm abs} = {1\over p}\left(1-|S_m|^2\right) = {1\over p}
\left(1- e^{-2\pi\mu}\right)\,, \quad |m-\beta|< \gamma\,.
\end{equation}

Now we come back to the radial solutions of Eq. (\ref{Re})
with arbitrary coefficients $a_m$ and $b_m$ and obtain the general
expression for the partial absorption cross sections. This permit us to
work out analytically the case of the total absorption that has been
considered phenomenologically in \cite{Audretsch98,Audretsch99}.

Subtracting the radial function of the incoming plane wave, modified by
the AB-phase factor \cite{Audretsch99}, from the solutions (\ref{rs2},
\ref{rs3}) we obtain the scattering amplitude (\ref{f}) with the phase
functions
\begin{equation} \label{incs1}
S_m = e^{i\pi (m-\mu)}\,, \quad |m-\beta|> \sqrt{1+\gamma^2}\,,
\end{equation}
\begin{equation} \label{incs2}
S_m = e^{i\pi (m-\mu)}\,{a_m \over b_m}\,,  \quad  \gamma<|m-\beta|< \sqrt{1+\gamma^2}
\end{equation}
\begin{equation} \label{incs3}
S_m = e^{i\pi m}\,e^{\pi \mu}\,{a_m \over b_m}\,,\quad |m-\beta|< \gamma\,.
\end{equation}
and the nonzero partial absorption cross sections
\begin{equation} \label{mod2}
\sigma_m^{\rm abs} = {1\over p} \left(1-{|a_m|^2 \over |b_m|^2}\right)\,,
\quad \gamma<|m-\beta|< \sqrt{1+\gamma^2}
\end{equation}
\begin{equation} \label{mod3}
\sigma_m^{\rm abs} = {1\over p} \left(1- e^{2\pi\mu}{|a_m|^2 \over
|b_m|^2}\right)\,, \quad |m-\beta|< \gamma\,.
\end{equation}
For these orbital modes the partial absorption cross
sections depend on the choice of the coefficients $a_m$ and $b_m.$
The different choice of these coefficients can be considered as the choice
of different models of absorption.

We turn now to the approach to the absorption problem that has been
proposed in \cite{Audretsch98,Audretsch99} for the case of the AB
scattering polarizable atoms by a charge wire of a finite radius. The notion
of an {\sl absorption radius} was introduced there, such that incident
particles with impact parameter $a=|m-\beta|/p<\rho_{\rm abs},$ which refers
to the kinetic angular momentum, will be absorbed. It determines an interval
$[-n_-,n_+]$ of the values of $m$ inside of which the condition of the total
absorption
\begin{equation}\label{totabs}
S_m = 0\,, \quad -n_- < m < n_+
\end{equation}
was imposed instead of the value $S_m=e^{i\pi m-\pi\mu}$ which followed
from Eq. (\ref{incs3}) at $b_m\,e^{-\pi\mu} = a_m\,e^{\pi\mu}.$ This
condition can be evidently derived analytically from Eq. (\ref{incs3})
assuming that the coefficient $a_m=0$ if {\sl the absorption radius} less
than {\sl the critical radius} defined by the values of $m$ for which
$\nu^2<0.$ The corresponding radial solution is the Hankel function
\begin{equation} \label{rsmin}
R_m (\rho) = {1\over 2}\,e^{i\pi m}\,e^{\pi\mu/2}\, H_{i\mu}^{(2)}(p\rho)
\end{equation}
which has the asymptotic behavior at $\rho\rightarrow\infty$ like
Eq. (\ref{asymR}) with $S_m=0.$ At small $\rho$ it contains both the
ingoing and outgoing waves, moreover, the outgoing wave is weakened so to
escape at infinity. In this case the partial scattering amplitudes are equal
to
$$
f_m = -{e^{-i{\pi\over 4}}\over p}\cos\pi\beta\,,\quad |m-\beta|<\gamma\,,
$$
and the partial absorption cross sections are equal to their maximal values
$$
\sigma_m^{\rm abs} = {1\over p}\,,\quad |m-\beta|<\gamma\,.
$$

For the intermediate case of Eq. (\ref{incs2}) the condition of the
total absorption $S_m=0$ is realized if $a_m=0.$ Then the corresponding
wave function reads
\begin{equation} \label{min}
R_m (\rho) = {1\over 2}\,e^{i\pi m}\,e^{-i\pi\mu/2}\, H_{\mu}^{(2)}(p\rho)\,.
\end{equation}
We note that absorption can happens for the pure AB scattering when the
attractive potential is absent ($\gamma=0$). Then, for the partial modes
with $|m-\beta|<1.$ i.e. for the values of $m=0$ and $m=1,$ it is possible
to get any value of the partial absorption cross sections up to its maximal
value $1/p.$ For the pure attractive potential ($\beta=0$) the number of
modes that can be absorbed, $|m|<\gamma,$ depends on the value of $\gamma.$

\section{Other singular attractive potentials}

We consider now the Schr\"{o}dinger equation Eq. (\ref{Re}) with the more
singular attractive potential
\begin{equation} \label{U1}
U(\rho) = -{\kappa^2\over\rho^4}\,.
\end{equation}
The Schr\"{o}dinger equation with this potential alone describes the
scattering of ions by polarization forces \cite{Vogt54}. We keep
the AB potential in addition to the potential (\ref{U1}). Then the radial
equation takes the form
\begin{equation}\label{Re1}
R''_m (\rho)+{1\over\rho}\,R'_m(\rho) {(m-\beta)^2\over\rho^2}\,R_m(\rho) +
{\lambda^2\over\rho^4}\,R_m(\rho)+p^2\,R_m(\rho) = 0
\end{equation}
and by the substitution  $\rho=\rho_0 e^x,\, \rho_0:=\sqrt{\lambda/p}$
can be reduced to the Mathieu equation of imaginary argument
\begin{equation}\label{Me}
R''_m (x)-(a-2q\cosh 2x)\,R_m(x)  = 0
\end{equation}
with the parameters $a:= (m-\beta)^2$ and $q:=p\lambda.$

Asymptotic behaviour of solutions of Eq. (\ref{Me}) at large $\rho$
(large positive $x$) is well known,
\begin{equation}\label{R12}
R_m^{(1, 2)}(\rho) \sim H_{m-\beta}^{(1), (2)}(p\rho)
\sim \sqrt{2\over\pi p\rho}\, e^{\pm i(p\rho-{\pi\over 2}|m-\beta|-
{\pi\over 4})}\,,\quad \rho \rightarrow \infty\,.
\end{equation}
The Mathieu functions Eq. (\ref{R12}) form a complete set and the total
solution of Eq. (\ref{Re1}) can be represented by their linear combination
\cite{Wannier54}
\begin{equation} \label{rs4}
R_m(\rho) = a_m\,R_m^{(1)}(\rho_0) + b_m\,R_m^{(1)}(\rho_0)\,.
\end{equation}
With these radial functions we obtain the following expression for the
phase function of Eq. (\ref{asymR})
\begin{equation}\label{incs4}
S_m = e^{i\pi (m-|m-\beta|)}\,{a_m\over b_m}\,.
\end{equation}
Now we can choose the coefficients $a_m$ and $b_m$ in accordance with the
different criteria discussed above. For this aim we have to consider the
asymptotic behaviour of solutions of Eq. (\ref{Me}) at small $\rho$
(large negative $x$). There are two other Mathieu functions which describe
the ingoing and outgoing waves nearby the singularity line $\rho=0,$
\begin{equation}\label{R34}
R_m^{(3,4)}(\rho) \sim H_{m-\beta}^{(1), (2)}({\lambda\over\rho})
\sim \sqrt{2\rho\over\pi\lambda}\, e^{\pm i({\lambda\over\rho}-
{\pi\over 2}|m-\beta|-{\pi\over 4})}\,,\quad \rho \rightarrow 0
\end{equation}
which also form a complete set. Now, to work out the elastic scattering,
we have to impose the self-adjointness condition on the Hamiltonian of
Eq. (\ref{Re1}) \cite{Case50}, taking the linear combination
\begin{equation}\label{Rel}
R_m^{(\rm el)}(\rho) = e^{-i\theta_m}\,R_m^{(3)}(\rho)+
e^{i\theta_m}\,R_m^{(4)}(\rho)\,.
\end{equation}
Using their connections with the Mathieu functions $R_m^{(1,2)}(\rho)$
we obtain the coefficients $a_m$ and $b_m$ and the phase function
(\ref{incs4}) of the elastic scattering. On contrary,
eliminating of the outgoing wave at $\rho= 0$ to describe absorption we have
to connect the solution $R_m^{(3)}(\rho)$ with the Mathieu functions
$R_m^{(1,2)}(\rho)$ to find another coefficients $a_m$ and $b_m.$ In this way the inelastic
scattering of ions by polarization forces has been considered and the
capture cross section has been evaluated in \cite{Vogt54}. At least,
to get the total absorption of given modes we have to put zero the
corresponding coefficients $a_m.$

The choice of the coefficients $a_m$ and $b_m$ depends on model of
absorption under consideration. For example, one can put $S_m =
S_m^{\rm el}$ for large $|m|\geq m_{\rm abs}$ (large impact parameters)
and $S_m=0$ for $|m|\leq m_{\rm abs}$ or to consider a intermediate regime
from the elastic scattering to the total absorption.
\medskip

Up to now we dealt with potentials that are singular on the line. The
similar results can be obtained for potentials with singularities on a
surface. For example, one can use the potential that is
singular on the cylinderic surface $\rho=\rho_0,$
$$
U(\rho) = -\kappa{e^{-|\rho-\rho_0|/l}\over (\rho-\rho_0)^n}\,\quad n\geq 2
$$
which will imitate absorption in a thin layer of width $\sim l$ nearby this
surface. Of cause, the analytical calculations in this case
are very cumbersome, if possible.

\section{Conclusion}

The main result of the paper is following: If partial wave functions
 of a quantum mechanical problem are not fixed by
the normalization conditions as this happens in the presence of singular
potentials, then the corresponding arbitrariness can be used to describe
not only elastic scattering but also inelastic one (absorption). By
different choice of partial radial wave functions various models of
absorption can be realized, beginning from almost elastic scattering
till maximal possible values of partial absorption coefficients.

We give up unsolved very important problem what type of physical situations
with regular potentials corresponds in a singular limit to different choice
of the partial wave functions describing elastic scattering or absorption.

\section*{Acknowledgments}

This work was supported by the Deutsche Forschungsgemeinschaft.

\end{document}